\documentclass[aps,prl,amsmath,amssymb,longbibliography,lengthcheck,showpacs,superscriptaddress,floatfix]{revtex4-1}
\usepackage{graphicx}
\usepackage{color}
\def\vec#1{\mbox{\boldmath $#1$}}

\begin{document}

\title{Spatial Distributions of Local Elastic Moduli Near the Jamming Transition}

\author{Hideyuki Mizuno}
\email{Hideyuki.Mizuno@dlr.de} 
\affiliation{Institut f\"{u}r Materialphysik im Weltraum, Deutsches Zentrum f\"{u}r Luft- und Raumfahrt (DLR), 51170 K\"{o}ln, Germany}

\author{Leonardo E.~Silbert}
\affiliation{Department of Physics, Southern Illinois University Carbondale, Carbondale, IL 62901 USA}

\author{Matthias Sperl}
\affiliation{Institut f\"{u}r Materialphysik im Weltraum, Deutsches Zentrum f\"{u}r Luft- und Raumfahrt (DLR), 51170 K\"{o}ln, Germany}

\date{\today}

\begin{abstract}
  Recent progress on studies of the nanoscale mechanical responses in
  disordered systems has highlighted a strong degree of heterogeneity in the
  elastic moduli.  In this contribution, using computer simulations, we study
  the elastic heterogeneities in athermal amorphous solids, composed of
  isotropic, static, sphere packings, near the jamming transition.  We employ
  techniques, based on linear response methods, that are amenable to
  experimentation.  We find that the local elastic moduli are randomly
  distributed in space and are described by Gaussian probability
  distributions, thereby lacking any significant spatial correlations, that
  persists all the way down to the transition point.  However, the shear
  modulus fluctuations grow as the jamming threshold is approached, which is
  characterized by a new power-law scaling.  Through this diverging behavior
  we are able to identify a characteristic length scale, associated with shear
  modulus heterogeneities, that distinguishes between bulk and local elastic
  responses.
\end{abstract}


\maketitle

When traditional, crystalline solids are linearly deformed, their elastic
responses are typically described by affine deformations~\cite{elastictheory}.
Contrary to this, disordered solids, such as thermal amorphous solids,
i.e.~glasses, disordered crystals~\cite{lowtem}, as well as athermal jammed
solids~\cite{Makse_2000}, exhibit strongly non-affine responses to elastic
deformations. This non-affine character becomes significantly apparent during
shear deformation~\cite{Tanguy_2002}. Under shear, constituent particles
undergo additional non-affine displacements~\cite{Maloney_2006}, leading to a
decrease in the shear modulus from a value predicted by the affine response
only~\cite{Tanguy_2002}. It is this non-affine character that dominates the
shear modulus on approach to the jamming transition, where a mechanically
stable solid loses rigidity~\cite{OHern_2003,Zaccone_2011}.

The appearance of non-affine response is closely related to elastic
heterogeneities~\cite{leonforte_2005}, especially spatially varying shear
moduli. Indeed, DiDonna and Lubensky~\cite{DiDonna_2005} proposed that
non-affine displacements of particles subject to shearing are driven by
randomly fluctuating local elastic moduli.  Amorphous solids reflect such
inhomogeneous behavior in their mechanical responses at the
nanoscale~\cite{Schirmacher_2015,Schirmacher2_2015,Hufnagel_2015}, as seen in
both computer simulations~\cite{Yoshimoto_2004} and
experiments~\cite{Wagner_2011}.  Manning and
co-workers~\cite{Manning_2011,chen_2011} identified soft spots as regions of
atypically large displacements in low-frequency, quasi-localized vibrational
modes. Particle rearrangements, activated by mechanical
load~\cite{Manning_2011,Tanguy_2010} and by thermal
energy~\cite{chen_2011,widmer_2008}, are therefore understood to be spatially
correlated with those soft spots, which can be linked to locally unstable
regions with negative shear moduli~\cite{Yoshimoto_2004}.  Furthermore,
Ellenbroek~\textit{et al.}~\cite{Ellenbroek_2006} demonstrated that the
elastic response of jammed packings to local forcing fluctuates over a length
scale $\ell_\ast$. Independently Lerner~\textit{et al.}~\cite{Lerner_2014}
showed that the local elasticity is governed by a different length $\ell_c$.
Recently Karimi and Maloney~\cite{Karimi_2015} reconciled these
differing views by considering the behaviors of longitudinal and
transverse components of elastic response.

Thus, it appears that spatial heterogeneities in local elastic moduli are
a key feature to understanding mechanical properties of disordered solids.  In
this contribution, we study the elastic heterogeneities in athermal jammed
solids close to the jamming transition.  Specifically, we address the
following points: (i) How are the local elastic moduli distributed in space?
(ii) How do those distributions evolve on approach to the jamming transition?
(iii) Is there a length scale over which the local elastic moduli fluctuate?
For athermal systems studied here, the packing fraction $\phi$ acts as a
control parameter that we use to systematically probe static packings of
varying rigidity.  We characterize rigidity by the distance, $\Delta \phi =
\phi - \phi_c$, from the transition point $\phi_c$, or equivalently
the packing pressure, $p$. The approach of $\phi_c$ from above ($p\rightarrow
0^{+}$) is governed by various power-law scalings with $\Delta \phi$ in
quantities including global elastic moduli~\cite{Makse_2000,OHern_2003,Ellenbroek_2006}.
In the following, we unveil new power-law scalings in the spatial fluctuations of elastic moduli.

Our numerical system consists of $N$ monodisperse, frictionless spheres of
diameter $\sigma$ and mass $m$, in three dimensional, periodic, cubic
simulation boxes~\cite{Silbert_2010}.  Particles interact via a finite-range,
purely repulsive potential; $V(r) = (\epsilon/a)(1-r/\sigma)^a$ for $r <
\sigma$, otherwise $V(r)=0$, where $r$ is the center-to-center separation
between two particles. Here, we show results only for Hertzian contacts,
$a=2.5$ \cite{jnote1}. Length, mass, and time are presented in units of
$\sigma$, $m$, and $\tau=(m\sigma^2 /\epsilon)^{1/2}$. We
prepared systems over several orders of magnitude in packing pressure,
$10^{-7} < p < 10^{-1}$, corresponding to $10^{-6} \lesssim \Delta \phi
\lesssim 10^{0}$.  Most of our results are for $N=1,000$, but we also show
data using $N=10,000$ to probe larger length scales.

The total elastic modulus (bulk, shear), $X (=K,G)$, is obtained as a sum of
the affine, $X_A$, and non-affine, $X_N$, components: $X = X_{A} -
X_{N}$~\cite{Lutsko_1989,Lemaitre_2006,Lutsko_1988,Wittmer_2013,Mizuno_2013}.
While $X_A$ can be thought of as the value predicted assuming particles follow
affine trajectories under an imposed deformation field, $X_N$ quantifies
deviations from this due to non-affine relaxations.  Yet, obtaining elastic
modulus information in fragile systems can be problematic, especially when
applying explicit deformation procedures.  Here, we implemented protocols
developed within linear response
theory~\cite{Lutsko_1989,Lemaitre_2006,Lutsko_1988,Wittmer_2013,Mizuno_2013},
which avoid explicit deformation practices thereby allowing us to probe
extremely close to the jamming transition.

Two protocols were employed that essentially sample the vibrational normal
modes of system: (i) The zero-temperature ($T=0$) protocol (restricted to
$N=1,000$) is formulated directly in terms of the dynamical
matrix~\cite{Lutsko_1989,Lemaitre_2006}.  (ii) The finite-temperature ($T>0$)
protocol (for both $N=1000$ and $N=10,000$), samples mode vibrations by
switching on a small temperature ($T=10^{-9}$ to $10^{-10}$) and thermally
agitating the system~\cite{Lutsko_1988,Wittmer_2013,Mizuno_2013}.  At these
temperatures and $p>10^{-5}$, particle displacements are $10^{-2}$ to
$10^{-4}$ $[\sigma]$, and both protocols return consistent values.  Technical
details of numerical procedure and formulation can be found in Supplemental Material~\cite{supplement}.
Here we highlight an important
aspect of these protocols.  Both procedures are accessible through current
experimental technologies at the colloidal and granular scales.  In
particular, advances in particle tracking and resolution allow precision
measurements of particle positions, used by covariance matrix analyses
methods~\cite{Ghosh_2010-2,Henkes_2012,Still_2014}, and the photo-elastic
technique for particle forces~\cite{Majmudar_2007}.

To extract local information, the simulation box was divided into
small subvolumes of size $w_x \times w_y \times w_z$, i.e.~coarse-graining
(CG) domains.  In each CG domain $m$, we computed the local modulus,
$X^{m}=K^m,G^m$, decomposed into their affine ($A$) and non-affine ($N$)
components.  We then calculated the probability distribution function
$P(X^m)$, from which the average $X$ and standard deviation $\delta X$ were
obtained~\cite{supplement}.  $\delta X=\delta X(p,w_x,w_y,w_z)$ depends on
both $p$ and the size of CG domain, and quantifies the extent of fluctuations,
whereas $X=X(p)$ corresponds to the global value, independent of $w_\alpha$
($\alpha=x,y,z$)~\cite{jnote13}.

\begin{figure}[t]
\centering
\includegraphics[width=0.48\textwidth]{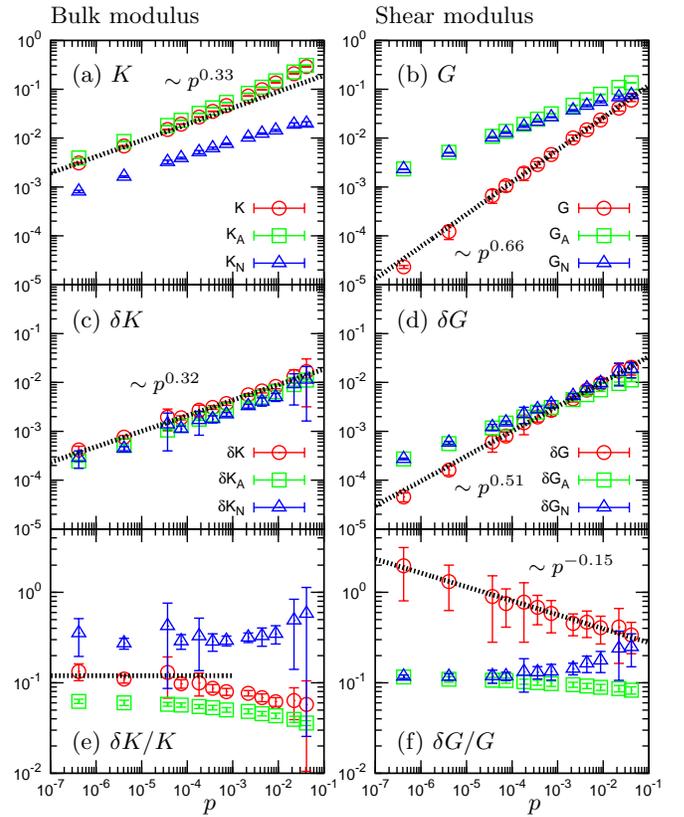}
\vspace*{-6mm}
\caption{\label{heterogeneity} (Color online) 
Elastic modulus dependence on packing pressure, $p$.
The average (global) total, affine ($A$), and non-affine ($N$) values $X, X_{A}, X_{N}$, and corresponding standard deviations $\delta X$ of the probability distribution $P(X^m)$, for $X^m=K^m$ (left panels) and $G^m$ (right panels).
The CG domain is cubic of linear size $w \simeq 3$.
Lines are power-law scalings with $p$.
The presented data were obtained using the $T=0$ protocol with $N=1,000$ and averaging over $10$ different realizations at each value of $p$.}
\end{figure}

Figure~\ref{heterogeneity} shows the dependence on pressure, $p$, of the
moduli and their corresponding fluctuations.  The global $X(p)$ are shown in
the top panels, Fig.~\ref{heterogeneity}(a), (b), indicating that our
technique is consistent with previous studies on similar
systems~\cite{OHern_2003,Ellenbroek_2006} that imposed explicit deformations.
Since the pressure scales as $p \sim V' \sim \Delta \phi^{a-1}$ ($\sim
\Delta \phi^{1.5}$ for $a=2.5$, Hertzian contacts), the scaling laws for $X$
normalized by the effective spring constant $k_\text{eff} \sim V'' \sim \Delta
\phi^{a-2}$~\cite{Vitelli_2010}, $X/k_\text{eff}$, are consistent with:
\begin{equation} \label{plmacro}
{K}/{k_\text{eff}} \sim \Delta \phi^0, \quad {G}/{k_\text{eff}} \sim \Delta \phi^{0.5}.
\end{equation}
The middle panels, Fig.~\ref{heterogeneity}(c), (d), show the absolute fluctuations, $\delta X(p,w_x,w_y,w_z)$, where the CG domain is cubic of linear size, $w_\alpha=w\simeq 3$, and from which we find,
\begin{equation}
\label{plfluctu}
{\delta K}/{k_\text{eff}} \sim \Delta \phi^0, \quad {\delta G}/{k_\text{eff}} \sim \Delta \phi^{0.27}.
\end{equation}
More importantly, the bottom panels, (e) and (f), present the fluctuation data
on a relative scale, $\delta X/X$, which gives the appropriate measure of the
degree of heterogeneity.  As $\Delta \phi \rightarrow 0$ ($p \rightarrow 0$),
$\delta K/K$ approaches a constant value, whereas relative fluctuations in the
shear modulus grow as
\begin{equation} \label{ghetero}
{\delta G}/{G} \sim \Delta \phi^{-\nu_G}, \quad \nu_G \simeq 0.5-0.27=0.23.
\end{equation}

We remark on two additional key features of Fig.~\ref{heterogeneity}.
Firstly, for the bulk modulus the affine and non-affine components are quite
distinct, such that the total bulk modulus is largely determined
by the affine part only.  Secondly, and in contrast to the above, the shear
modulus components remain close in value, so the scaling for total shear
modulus is controlled by the gradual cancellation of affine and non-affine
contributions.

\begin{figure}[t]
\centering
\includegraphics[width=0.48\textwidth]{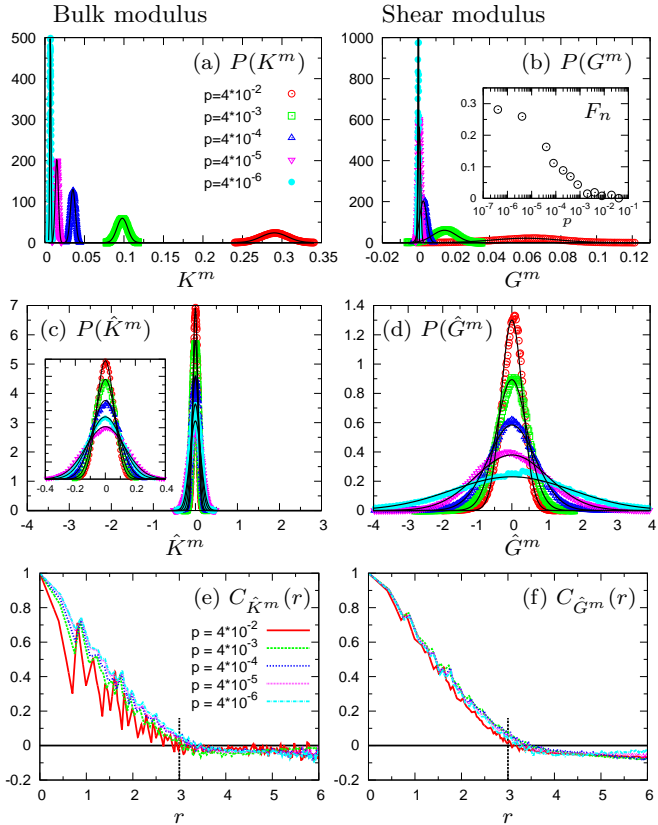}
\vspace*{-6mm}
\caption{\label{distribution} (Color online)
Probability distributions of (a) $K^m$ and (b) $G^m$, and their relative fluctuations, (c) $\hat{K}^m = (K^m-K)/K$ and (d) $\hat{G}^m = (G^m-G)/G$, for the range of $p$ indicated in the legend of panel (a).
Spatial correlation functions $C_{\hat{X}^m}(r)$ (defined in main text) for (e) $\hat{K}^m$ and (f) $\hat{G}^m$.
The inset to (b) shows the fraction of negative $G^m$ regions, $F_n=\int_{G^m<0} P(G^m) dG^m$, as a function of $p$.
A close-up of $P(\hat{K}^m)$ is shown in the inset to (c).
In (a)-(d), solid lines indicate Gaussians.
In (e),(f), vertical lines indicate the CG length, $r = w_\alpha = w \simeq 3$.
Data were obtained using the $T=0$ protocol for $N=1,000$.}
\end{figure}

\begin{figure}[t]
\centering
\includegraphics[width=0.48\textwidth]{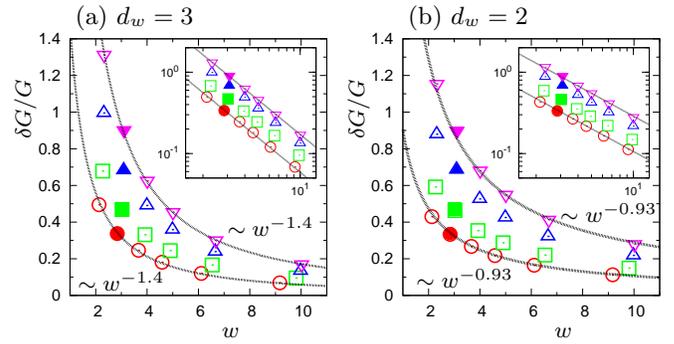}
\vspace*{-6mm}
\caption{\label{whetero} (Color online)
Dependence of $\delta G/G$ on the CG length $w$ for (a) $d_w=3$ and (b) $d_w=2$, as discussed in main text.
Same symbols used as key in Fig.~\ref{distribution}(a).
Closed symbols are data using $w_\alpha =w \simeq 3$, i.e. same data as shown in Fig.~\ref{heterogeneity}(f).
Lines are power-law scalings, (a) $\delta G/G \sim w^{-1.4}$ and (b) $\sim w^{-0.93}$, consistent with $\delta G/G \sim w^{-d_w/2}$.
Insets: same plots on log-log scales.
For $w<6$ data were obtained by the $T=0$ protocol with $N=1,000$, and for $w>6$ the $T>0$ protocol with $N=10,000$.}
\end{figure}

\begin{figure*}[t]
\centering
\includegraphics[width=0.88\textwidth]{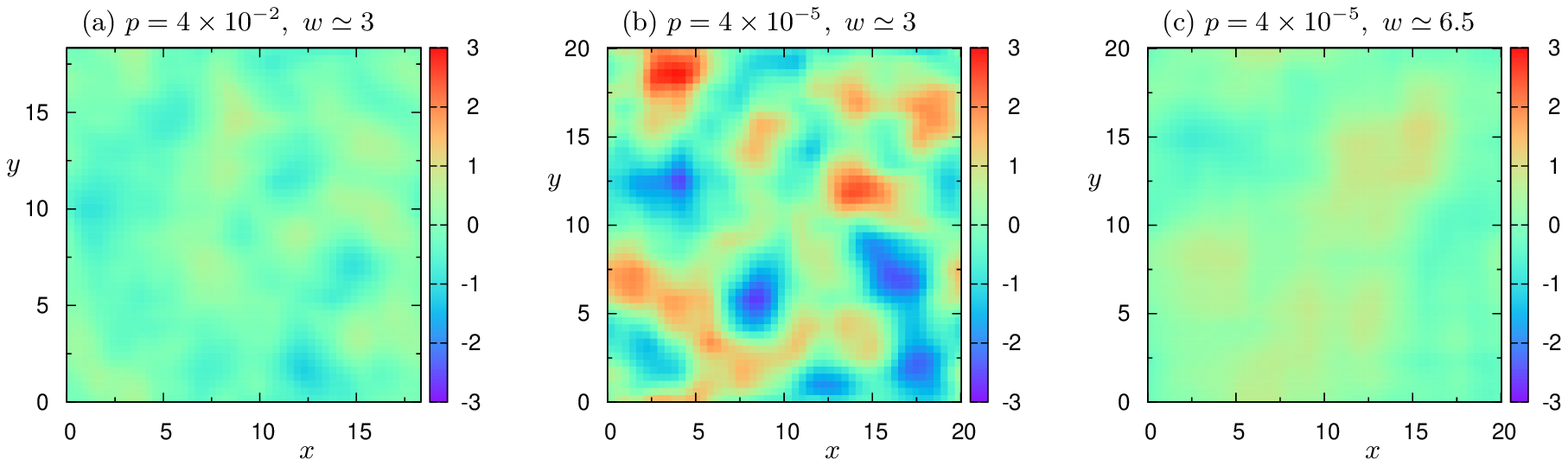}
\vspace*{-2.5mm}
\caption{\label{map} (Color online)
Spatial maps of local shear modulus fluctuations, $\hat{G}^m=(G^m-G)/G$, within a fixed $x$-$y$ layer, for $d_w=3$.
(a) Large pressure $p=4\times 10^{-2}$ and small CG length $w \simeq 3$, (b) small $p=4\times 10^{-5}$ and small $w \simeq 3$, and (c) small $p=4\times 10^{-5}$ and large $w \simeq 6.5$.
Data were obtained using the $T>0$ protocol with $N=10,000$.
Additional snapshots shown in Supplemental Material~\cite{supplement}.}
\end{figure*}

We now turn to a more explicit view of the spatial distributions of $K^m$ and
$G^m$.  Figure~\ref{distribution} presents the probability distributions
$P({K}^m)$ in (a) and $P({G}^m)$ in (b).  We find that all the $P(X^{m})$ are
well-characterized as Gaussian over the entire pressure range, even down to
the jamming point~\cite{jnote11}. But notice that although all the $K^{m} >
0$, $G^{m}$ can contain negative values.  The fraction of these negative shear
modulus zones, $F_{n} = \int_{G^{m}<0} P(G^{m})dG^{m}$, is shown in the inset
to Fig.~\ref{distribution}(b). $F_{n}$ grows as $p \to 0$, suggesting that
there is a $1:1$ ratio of stable and unstable regions~\cite{jnote15} as the
system becomes fragile~\cite{cates1998}.
Note the fact that our data appear to level off at the lowest pressure is likely a system size effect~\cite{Goodrich_2012}.
In Fig.~\ref{distribution}(c), (d), we plot $P(\hat{K}^m)$ and $P(\hat{G}^m)$
of the fluctuations relative to global value, $\hat{X}^m = (X^m-X)/X$.
$P(\hat{G}^m)$ broadens significantly as $p$ decreases, which is
quantitatively demonstrated by $\delta G/G$ in
Fig.~\ref{heterogeneity}(f)~\cite{jnote5}, whereas variations in
$P(\hat{K}^m)$ are rather small and insensitive to $p$, consistent with
$\delta K/K$ in Fig.~\ref{heterogeneity}(e).

In an effort to directly detect a correlation length associated with these
fluctuations, the bottom panels of Fig.~\ref{distribution}(e), (f) show the
fluctuation spatial correlation function, $C_{\hat{X}^m}(r) = { \left<
    \hat{X}^m(\vec{r}) \hat{X}^m(\vec{0}) \right> }/{ \left<
    \hat{X}^m(\vec{0}) \hat{X}^m(\vec{0}) \right>}$, where we explicitly
represent $\hat{X}^m$ as a function of position $\vec{r}$, and $\left<
\right>$ denotes a spatial average.  Both the $C_{\hat{X}^m}(r)$ decay with
the CG length $r = w \simeq 3$~\cite{jnote9}, indicating that $K^m$ and $G^m$
fluctuate randomly in space without any apparent correlation, which persists
all the way down to the transition point.  Thermal
glasses~\cite{Yoshimoto_2004,Mizuno_2013,Tsamados_2009,Mizuno2_2013,Mizuno_2014}
and disordered crystals~\cite{Mizuno2_2013,Mizuno_2014} similarly exhibit
random distributions in their local moduli that are Gaussian.

An alternative view to determining a possible characteristic length is
through the dependence of fluctuations, $\delta X/X$, on the size of CG
domain, $w_x \times w_y \times w_z$.  We considered three different ways to
change the CG domain: Vary, (i) $w_x,w_y,w_z$ equally, so that
$w_x=w_y=w_z=w$, (ii) $w_x,w_y$ as $w_x=w_y=w$, keeping fixed $w_z \simeq 3$,
(iii) only $w_x$ as $w_x=w$, keeping fixed $w_y=w_z \simeq 3$.  In (i), the CG
domain is always cubic, whereas it becomes rectangular parallelepiped in (ii),
(iii).  We define the dimension $d_w$ of CG domain; $d_w=3,\ 2,\ 1$ for
(i), (ii), (iii).  As we have seen so far, $X^m$ is a random
variable, following a Gaussian $P(X^m)$.  Thus, within the framework of a sum
of random variables~\cite{jnote6}, we obtain the scaling law with respect to
CG length $w$:
\begin{equation}
\label{dwscaling}
{\delta X}/{X} \sim w^{-d_w/2}.
\end{equation}
Figure~\ref{whetero} shows the $w$-dependence of $\delta G/G$ at several different $p$, for $d_w=3$ in (a) and $d_w=2$ in (b) (see~\cite{supplement} for $d_w=1$).
For all pressures, $\delta G/G \sim w^{-1.4}$, $\sim w^{-0.93}$, $\sim w^{-0.47}$ for $d_w=3$, $2$, $1$, respectively, which all confirm Eq.~(\ref{dwscaling}).
We obtained the same result in $\delta K/K$.
The same power-law dependence on $w$ has been reported for glasses, with exponent $1$ in $d_w=2$~\cite{Tsamados_2009} and $1.5$ in $d_w=3$~\cite{Mizuno_2013}.

Combining the scaling results for $\delta G/G$ (Eqs.~(\ref{ghetero}) and~(\ref{dwscaling})), expresses that relative fluctuations in shear modulus are suppressed over sufficiently large $w$.
This supports the existence of a characteristic length, $\xi_G$, above which fluctuations become negligible.
Specifically, we define $\xi_G$ as $w$ at which we see a fixed value, $\alpha_0$, of $\delta G/G$ for all $p$ or $\Delta \phi$, i.e. we determine $\xi_G$ as $\delta G/G = \alpha_0 (w/\xi_G)^{-d_w/2}$, which gives~\cite{jnote7a,jnote7b}
\begin{equation} \label{length}
\xi_G \sim \Delta \phi^{-\nu_{\xi}}, \quad \nu_{\xi} = \nu_G/(d_w/2).
\end{equation}
The idea of the length $\xi_G$ associated with growing $\delta G/G$ is best visualized in Fig.~\ref{map}, which shows the local fluctuations of shear modulus (for $d_w=3$) as follows: Panels (a) and (b) of Fig.~\ref{map} compare modulus maps of $\hat{G}^m=(G^{m}-G)/G$ for a slice through two packings at two different $p$, at the same $w\ \simeq 3$.
In relation to Fig.~\ref{whetero}(a) ($d_w=3$), these two points lie at different values of $\delta G/G$ along a vertical line at $w \simeq 3$, that intersect
the respective $p$ curves.
At this value of $w$, the two systems appear very different.
Far from $\phi_{c}$, Fig.~\ref{map}(a) ($p = 4\times 10^{-2}$), the system appears quite uniform, and fluctuations are suppressed.
Whereas, close to $\phi_{c}$, Fig.~\ref{map}(b) ($p =4\times 10^{-5}$), we observe large-scale, spatial fluctuations.
For the system closer to $\phi_c$ (small $p$), fluctuations become suppressed at the larger $w=6.5$ (Fig.~\ref{map}(c)), so that the map resembles more compressed system at the smaller value of $w$.
This corresponds to drawing a horizontal line across Fig.~\ref{whetero}(a) at the same value of $\delta G/G$ connecting the two curves at different $p$.

In conclusion, we found that the differeces between bulk and shear
moduli fluctuations, as the jamming point is approached, are caused by the
non-affine components.  Relative fluctuations in the bulk modulus become
insensitive to packing pressure as $\Delta\phi \rightarrow 0$.  Whereas, shear
modulus fluctuations increase as, $\delta G/G \sim \Delta \phi^{-\nu_G}$,
which leads to the identification of a lengthscale, $\xi_G \sim \Delta
\phi^{-\nu_\xi}$.  For CG dimension, $d_{w}=3$, $\nu_{\xi} \approx 0.16$, a
value distinct from any previous study
\cite{Ellenbroek_2006,Lerner_2014,Karimi_2015,Vitelli_2010,Goodrich_2013,Schoenholz_2013,Ikeda_2013}.
$\xi_G$ corresponds to a scale above which the elastic properties coincide
with those of the bulk system, while below, the local mechanical properties
deviate from macroscopic behavior.  It has been proposed that a continuum
elastic description breaks down below a scale, $\ell_c \sim \Delta
\phi^{-1/4}$ in two dimensions~\cite{Lerner_2014}, consistent with our $\xi_G$
for $d_w=2$, and can derive from the transverse component of elastic
response~\cite{Karimi_2015}, which are controlled by shear modulus
fluctuations.

At the same time, however, we also found that the local elastic moduli
randomly fluctuate without any apparent correlations.  This feature
seems to be general for a wide class of disordered materials, thus further
promoting the idea that granular-like particle systems present a model
state for examining mechanical properties of disordered materials.
Curiously, the randomness in local moduli persists down to the transition
point and is different from the distribution of contact forces, which becomes
more exponential closer to $\phi_c$~\cite{Makse_2000} and is therefore more
suggestive of spatial correlations. Such random fluctuations in the moduli
may come from the coarse-graining procedure and/or the random distribution of
particle contacts, which is a topic for future investigation.

\begin{acknowledgments}
We acknowledge useful discussions with F.~Varnik, J.-L.~Barrat, S.~Mossa, W.~Schirmacher, K.~Saitoh, A.~Ikeda, C.~E.~Maloney, and A.~Zaccone.
H.M. acknowledges support from DAAD (German Academic Exchange Service).
L.E.S. gratefully acknowledges the support of the German Science Foundation DFG during a hospitable stay at the DLR under the grant FG1394.
M.S. acknowledges that during the stay at KITP, this research was supported in part by the National Science Foundation under Grant No. NSF PHY11-25915, as well as DFG FG1394.
\end{acknowledgments}

\bibliographystyle{apsrev4-1}
\bibliography{reference}

\begin{thebibliography}{53}
\bibitem [{jno({\natexlab{a}})}]{jnote1}%
  \BibitemOpen
  \href@noop {} {}\bibinfo {note} {Most of our results also hold for the
  one-sided harmonic potential, $a = 2$, although some differences arise, that
  will be explained in follow-up work.}\BibitemShut {Stop}%
\bibitem [{jno({\natexlab{e}})}]{jnote5}%
  \BibitemOpen
  \href@noop {} {}\bibinfo {note} {Here notice that the average value of
  $P(\hat{X}^m)$ is $\int \hat{X}^m P(\hat{X}^m) d\hat{X}^m =0$, and the
  standard deviation is $\sqrt{ \int (\hat{X}^m -0)^2 P(\hat{X}^m) d\hat{X}^m }
  = \delta X/X$.}\BibitemShut {Stop}%
\bibitem [{jno({\natexlab{f}})}]{jnote9}%
  \BibitemOpen
  \href@noop {} {}\bibinfo {note} {When $r\le w$, two CG domains to define
  local moduli are overlapped. The finite correlation of $C_{\hat{X}^m}(r)$
  within $r \le w$ comes from this overlapp, and the linear decay with $r$ is a
  consequence that the overlapp is reduced. At $r=w$, the correlation vanishes
  since the overlapp becomes zero.}\BibitemShut {Stop}%
\bibitem [{jno({\natexlab{g}})}]{jnote6}%
  \BibitemOpen
  \href@noop {} {}\bibinfo {note} {Let us start with the random variable
  $X_0^m$ at the CG length $w_0$, which follows a Gaussian distribution with
  the average $X$ and the standard deviation $\delta X_0$. When we increase the
  CG length as $w= n w_0$ ($n$ is an integer), $X^m$ can be written as a sum of
  $X_0^m$, i.e. $X^m = (1/M) \sum_{n=m_1}^{m_M} X_0^n = (1/M) (X^{m_1}_0 +
  X^{m_2}_0 + \dots + X^{m_M}_0)$, where $M=n^{d_w}$. Then, $X^m$ is also a
  random variable and likewise follows a Gaussian with the same average $X$ but
  smaller $\delta X = \delta X_0 \times M^{-1/2} = \delta X_0 \times
  (w/w_0)^{-d_w/2}$.}\BibitemShut {Stop}%
\bibitem [{jno({\natexlab{h}})}]{jnote7a}%
  \BibitemOpen
  \href@noop {} {}\bibinfo {note} {Here we note that the value $\xi_G$ itself
  depends on $\alpha_0$, but the exponent, $\nu_\xi = \nu_G/(d_w/2)$, does
  not.}\BibitemShut {Stop}%
\bibitem [{jno({\natexlab{i}})}]{jnote7b}%
  \BibitemOpen
  \href@noop {} {}\bibinfo {note} {We also define a length $\xi_K$ associated
  with $\delta K/K$, which converges to a constant value as $\Delta \phi
  \rightarrow 0$.}\BibitemShut {Stop}%
\end{thebibliography}

\end{document}